\documentclass[useAMS,usegraphicx]{mn2e}
\usepackage{rotate}
\usepackage{times}
\newif\ifAMStwofonts
\AMStwofontstrue
\def\gsim{\mathrel{\hbox{\rlap{\hbox{\lower4pt\hbox{$\sim$}}}\hbox{$>$}}}}
\def\lsim{\mathrel{\hbox{\rlap{\hbox{\lower4pt\hbox{$\sim$}}}\hbox{$<$}}}}

\def\Msun{M$_{\odot}$}

\def\exosat{{\it EXOSAT}}
\def\xmm{{\it XMM-Newton}}

\def\xmm{{\it XMM-Newton}}

\def\et{{et al.\ }}
\def\mcg{{MCG--6-30-15}}
\def\mrk{{Mrk~766}}

\def\Msun{\hbox{$\rm\thinspace M_{\odot}$}}

\title[The power spectrum of \mrk ]
      {The high frequency power spectrum of Markarian 766}

\author[Vaughan \& Fabian]
       {S. Vaughan\thanks{E-mail: sav@ast.cam.ac.uk} and
        A. C. Fabian \\
Institute of Astronomy, University of Cambridge, Madingley Road, Cambridge CB3 0HA\\
}
\date{Accepted 10/1/2003; submitted 17/12/2002; in original form 28/11/2002}
\pagerange{\pageref{firstpage}--\pageref{lastpage}}
\pubyear{2002}
\begin{document}
\maketitle
\label{firstpage}

\begin{abstract}
An analysis is presented of the power spectrum of X-ray variability
of the bright Seyfert 1 galaxy \mrk\ as observed by \xmm.  Over the
0.2--10~keV energy range the power spectral density (PSD) is
well-represented by a power-law with a slope of $\alpha_{\rm low}
\approx 1$ at low frequencies, breaking to a slope of $\alpha_{\rm
hi}=2.8_{-0.4}^{+0.2}$ at a frequency $f_{\rm br} \approx 5 \times
10^{-4}$~Hz.  As has been noted before this broken power-law PSD shape
is similar to  that observed in the Galactic black hole candidate
Cygnus X-1.  If it is assumed that \mrk\ shows a power spectrum
similar in form to that of Cyg X-1, and that the break timescale
scales linearly  with black hole mass, then the mass of the black hole
in \mrk\ is inferred to be $\lsim 5 \times 10^{5}$~\Msun. This rather
low mass would mean \mrk\ radiates above the Eddington limit.  The
coherence between different energy bands is significantly below unity
implying that variations in the different energy bands are rather
poorly correlated.  The low coherence can be explained in the
framework of standard Comptonisation models if the properties of the
Comptonising medium are rapidly variable or if there are several
distinct emission sites.
\end{abstract}


\section{Introduction}

X-ray variability appears to be ubiquitous in Active Galactic Nuclei
(AGN). 
The rapid and seemingly random fluctuations
in the X-ray luminosity of Seyfert galaxies provided early support for
the standard black hole/accretion disc model (e.g. Rees 1984) by
implying compact emission regions and high luminosity densities
(Fabian 1979; M$^{\rm c}$Hardy 1989).

The long ($\sim 3$ day), uninterrupted observations possible with
\exosat\ allowed the power spectral density (PSD) of Seyfert 1
galaxies to be measured for the first time (Lawrence \& Papadakis
1993; Green, M$^{\rm c}$Hardy \& Lehto 1993).   The PSD represents the
average of the (squared) amplitude of variations as a function of
temporal frequency (see van der Klis 1989).  The \exosat\ observations
showed that the PSDs of Seyfert 1s above $\sim 10^{-5}$~Hz could be
approximated by a power-law  ($\mathcal{P}(f) \propto f^{-\alpha}$
where $\mathcal{P}(f)$ is the power at frequency $f$ and $\alpha$ is
the PSD slope) rising steeply at low frequencies with a slope $\alpha
\gsim 1$ (so-called ``red noise''  spectra; Press 1978). It was noted
early on (e.g. Lawrence \et 1987; M$^{\rm c}$Hardy 1989) that this red
noise variability of Seyferts is similar to that observed in Galactic
Black Hole Candidates (GBHCs; van der Klis 1995), perhaps suggesting
that the same physical processes operate in these sources that differ
in black hole mass by factors of $\gsim 10^5$.

The steep slopes of the \exosat\ PSDs mean that at
lower frequencies the PSD must flatten in order for the integrated
power to converge. Indeed, breaks in the PSDs of Seyfert galaxies
have recently been detected (Edelson \& Nandra 1999; Uttley, M$^{\rm
c}$Hardy \& Papadakis 2002; Vaughan, Fabian \& Nandra 2003; Markowitz
\et 2003). The position of these breaks represent ``characteristic
timescales'' in the aperiodic variability of Seyfert 1s, and
interestingly they appear to scale linearly with the  mass of the
central black hole (Markowitz \et 2003). The PSDs of GBHCs usually
show at least two breaks (Nowak \et 1999; Lin \et 2000), and it is
becoming increasingly apparent that the PSDs of Seyfert galaxies (and
their timing properties in general) are very similar to those of GBHCs
except shifted to longer timescales according to the black hole mass
(Vaughan \et 2003; Markowitz \et 2003).

This letter presents an analysis of the X-ray continuum variability of
the bright Seyfert 1 galaxy \mrk\ (aka. NGC 4253; $z=0.012929$) using
a long \xmm\ observation.   \xmm\ offers the possibility to improve
upon the high frequency PSDs provided by \exosat,  having as it does a
similarly long orbit but greatly  increased throughput and a wider
bandpass. The rest of this paper is organised as
follows. Section~\ref{sect:data} describes the basic data reduction
procedures, Section~\ref{sect:psd} details the results of the PSD
analysis and Section~\ref{sect:cross} describes the cross spectral
results. Finally the implications of these are discussed in
section~\ref{sect:disco}.

\section{Data reduction}
\label{sect:data}

\begin{figure*}
\rotatebox{270}{
\scalebox{0.65}{
{\includegraphics{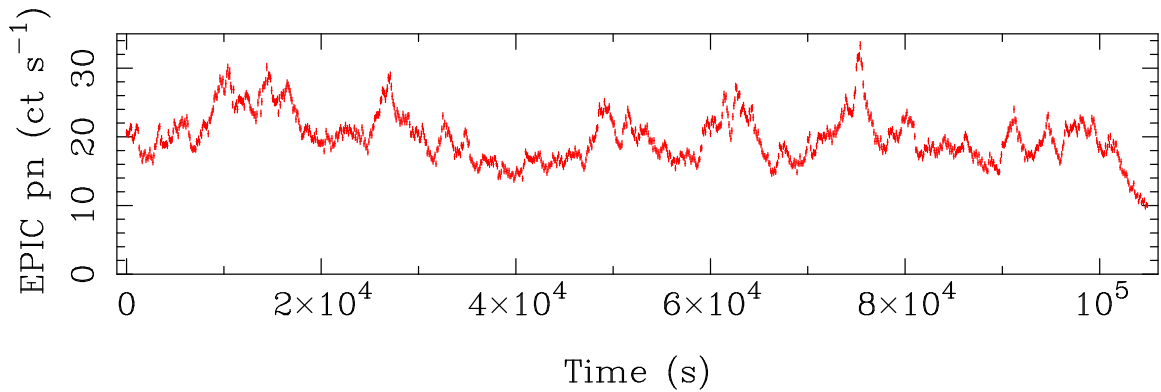}}}}
\caption{
Full-band (0.2--10.0~keV) EPIC pn light curve in 100~s bins. 
\label{fig:lc}}
\end{figure*}

\mrk\ was observed by \xmm\ (Jansen \et 2001) over the period 2001 May
20 -- 2001 May 21 (rev. 265), during which all instruments were
operating nominally. The present analysis is restricted to the data
from the European Photon Imaging Cameras (EPIC).  Other aspects of
this observation are discussed in Mason \et (2002).  The EPIC pn
camera (Str\"{u}der \et 2001) was operated in small-window mode, as
was the MOS2 camera (Turner \et 2001), and MOS1 was operated in timing
mode. Here only data from pn are used as these have the highest
signal-to-noise ratio and are free from photon pile-up.

Extraction of science products from the Observation Data Files (ODFs)
followed standard procedures using the \xmm\ Science Analysis System
v5.3.3 (SAS).  The EPIC data were processed using the standard SAS
processing chains, source data were extracted from a circular region
of radius 35 arcsec from the processed pn image.  Only events
corresponding to patterns 0--4 (single and double pixel events) were
used for the pn analysis. Background events were extracted from
regions in the small window least effected by source photons.  These
showed the background to be relatively low and stable throughout the
first 105~ksec of the observation.  During the final $\sim 15$~ksec
the background rate increased dramatically.  The much stronger (and
highly variable) background level during this part of the observation
contributed a significant amount of non-source variability, and so
only the first 105~ksec of the light curve are used for  the
variability analysis presented here.

Light curves were extracted from the EPIC pn data in four different
energy bands: 0.2--10.0~keV (full band), 0.2-0.7~keV (soft band),
0.7-2.0~keV (medium band) and 2.0-10.0~keV (hard band). These were
corrected for telemetry drop outs (less than 1 per cent of the total
time), background subtracted and binned to 100~s time resolution. The
errors on the light curves were calculated by propagating the Poisson
noise.  The light curves were not corrected for the $\sim 71$ per cent
``live time'' of the pn camera (Str\"{u}der \et 2001), which is only a
scaling factor.  The full band light curve is shown in
Fig.~\ref{fig:lc}.  The average source (background subtracted) and
background count rates are shown in Table~\ref{tab:basic} along with
the fractional excess rms variability amplitude of the source ($F_{\rm
var}$; Edelson \et 2002) in each energy band.

\begin{table}
\centering
\caption{
Basic properties of the light curves. 
Mean source and background count rates and fractional excess rms
variability amplitudes ($F_{\rm var}$) in each energy band.
}
 \begin{center}
\begin{tabular}{lccc}                
\hline
Band & source (ct s$^{-1}$) & background (ct s$^{-1}$) & $F_{\rm var}$ (\%) \\
\hline
Full   & $19.9$ & $0.3$  & $17.6\pm0.4$ \\
Soft   & $11.2$ & $0.2$  & $17.8\pm0.4$ \\
Medium & $6.68$  & $0.07$  & $19.0\pm0.4$ \\
Hard   & $2.03$  & $0.05$  & $17.8\pm0.4$ \\
\hline
\end{tabular}
\end{center}
\label{tab:basic}
\end{table}

\section{Power spectral analysis}
\label{sect:psd}

\begin{figure}
\rotatebox{270}{
\resizebox{!}{\columnwidth}
{
{\includegraphics{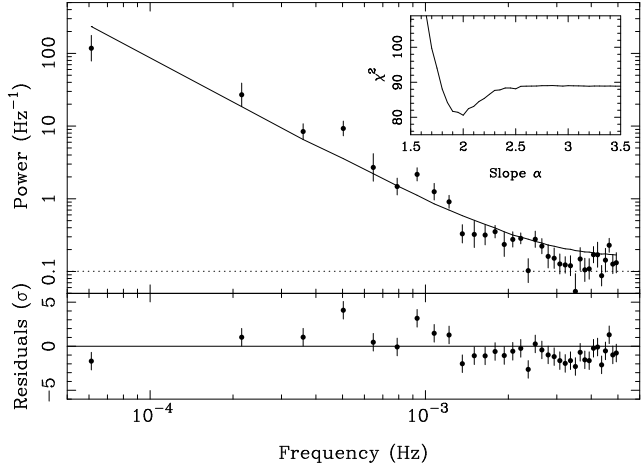}}}}
\caption{
Top panel: binned periodogram of the full band (0.2--10~keV) light curve
(data points) and best-fitting single power-law model (solid
line). The dotted line marks the expected Poisson noise level.
Bottom panel: residuals from the power-law fit. 
Inset panel: variation in $\chi^2$ with power-law slope.
\label{fig:psd1}
}
\end{figure}

The analysis presented here followed very closely that applied by
Vaughan \et (2003) to \xmm\ light curves of the bright Seyfert 1
galaxy \mcg.  The periodogram of the contiguous (100~s binned) light
curve was calculated using the standard Discrete Fourier Transform
(DFT). The periodogram was then binned using the method of Papadakis
\& Lawrence (1993), using $N=15$ periodogram points per bin, to
produce a consistent PSD estimate with Gaussian error
bars
%\footnote{The analysis was also performed using $N=20$ binning and
%the results were entirely consistent.}.  
The periodogram was
normalised to ``(rms/mean)$^{2}$ Hz$^{-1}$'' units as defined by van der Klis
(1997). The binned periodogram is shown in Fig.~\ref{fig:psd1} and
appears steep at low frequencies (due to the red noise variability of
the source) and flattens at high frequencies due to the variability
introduced by the Poisson noise (which has a flat, ``white'' power
spectrum).

The periodogram was fitted using the Monte Carlo procedure
described in  Vaughan \et (2003) (similar to that discussed
by Uttley \et 2002) in order to derive parameters of the PSD of \mrk.
For each trial PSD model 500 simulated light
curves were generated (using the algorithm of Timmer \& K\"{o}nig
1995).  The simulated light curves were generated from a PSD model
that extended down to much lower frequencies than the measured
periodogram, and each simulated light curve was a factor $\sim 50$
longer than the observed light curve. This allows for variability
power on timescales longer than the observed light curve and thereby
accounts for any effect this might have on the observed periodogram
(``red noise leak'' -- see Uttley \et 2002 and Vaughan \et
2003).  A section of each long simulated light curve was re-sampled to
match the  sampling of the observed light curve.  Periodograms were
then calculated for each of the 500 simulated datasets (and binned
exactly as the real data were) and the 500 binned periodograms
averaged to produce an average periodogram. This average represents
the PSD model after being distorted by the light curve sampling
(``folded'' through the response of the observation). The constant
Poisson noise component is then added (at the level expected for
Poisson noise) and the folded model is compared to the data by
measuring the  $\chi^2$ of the fit. Further details of the procedure
are given in Vaughan \et (2003).

Fitting the data with a power-law PSD model gave a poor fit to the
data ($\chi^2=80.8$ for $33$ degrees for freedom $dof$, rejected at
$>99.9$ per cent confidence) with a best-fitting power-law slope of
$\alpha=1.95$. The steep slope seen from these \xmm\ data requires
that the PSD flatten at lower frequencies (as observed in other
Seyfert 1 galaxies; Uttley \et 2002; Vaughan \et 2003; Markowitz \et
2003) for the total integrated power to be finite.  The poor fit of
the single power-law model further suggests that the power spectral
break may occur within  (or close to) the frequency range probed by
this \xmm\ observation.

\begin{figure}
\rotatebox{270}{
\resizebox{!}{\columnwidth}
{
{\includegraphics{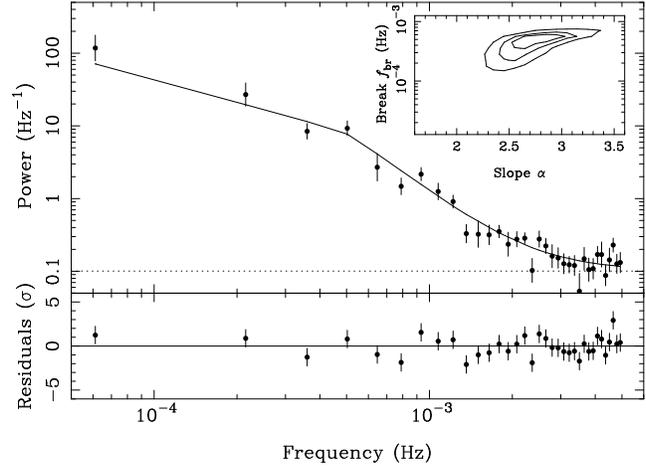}}}}
\caption{
Top panel: binned periodogram of the full band (0.2--10~keV) light curve
(data points) and best-fitting broken power-law model (solid
line). Bottom panel: residuals from the broken power-law fit. 
Inset panel: 68.3, 90 and 99 per cent confidence contours for the
model parameters (break frequency and power-law slope above the break).
\label{fig:psd2}
}
\end{figure}

In order to test this hypothesis a broken power-law PSD model was
fitted to the data.  The PSD was assumed to break from a slope of
$\alpha_{\rm low}=1$ at low frequencies to a steeper slope.
The free parameters of the model were the PSD normalisation, the break
frequency ($f_{\rm br}$) and the high frequency slope ($\alpha_{\rm
hi}$).  As shown in Fig.~\ref{fig:psd2}, this model provided a
reasonable fit to the data ($\chi^2 = 43.2 / 32 ~ dof$ with a
rejection probability of $91.5$ per cent), much better than the
unbroken power-law ($\Delta \chi^2 = 37.6$ improvement for one
additional parameter). The slightly high rejection probability may be
indicating  that a broken power-law is only an approximation to the
true underlying PSD.  The best-fitting parameters are as follows:
$\alpha_{\rm hi} = 2.8_{-0.4}^{+0.2}$ and break frequency of $f_{\rm
br} = 5_{-3}^{+1} \times 10^{-4}$~Hz. (The errors represent $90$ per
cent confidence limits calculated using a  $\Delta \chi^{2}=2.7$
criterion.)

The PSD slope below the break is not well-constrained with these \xmm\
data. An alternative model, in which the power-law breaks from
$\alpha_{\rm low}=0$ to a steeper slope was also compared to the
data. The best-fitting high frequency slope was $\alpha_{\rm
hi}=2.4_{-0.2}^{+0.3}$ and the break frequency was $1.6 \pm 0.8 \times
10^{-4}$~Hz, giving $\chi^2 = 48.1/32 ~ dof$ (rejected at $96.6$ per
cent), slightly worse than the model assuming a break from
$\alpha_{\rm low}=1$. Thus it is not possible to constrain in detail
the shape of break in \mrk, but  the broken power-law models did
provide a significant improvement over the single power-law model.  As
a low frequency slope of $1$ fitted the data slightly better, and
seems more likely in other Seyfert galaxies (Uttley \et 2002;
Markowitz \et 2003), this model is used throughout the rest of this
paper to describe the PSD of \mrk.

Energy dependence of the high frequency PSD, with the PSD slope being
flatter at higher energies, has previously been seen in the Seyfert 1
galaxies NGC 7469 (Nandra \& Papadakis 2001), \mcg\ (Vaughan \et 2003)
and NGC 4051 (Papadakis \& Lawrence 1995; M$^{\rm c}$Hardy \et 2003).
In order to test for any such variation in the PSD as a function of
energy the above analysis was repeated on the light curves extracted
in the three energy sub-bands. The results are shown in
Table~\ref{tab:e-psd}.  
The high frequency slopes are formally
consistent with one another although the hardest band does show the
flattest best-fitting slope (this is also the case when the data
were fitted with the PSD model assuming $\alpha_{\rm low}=0$ below the break).
This hints that \mrk\ may show the energy dependence seen in other
Seyferts, and in particular adds to the evidence provided by \xmm\ (see Vaughan \et 2003
and M$^{\rm c}$Hardy \et 2003) that the PSDs of Seyferts are energy-dependent
even at soft X-ray energies (the energy dependence in NGC~7469 could only be
examined above 2~keV, the same is true for most GBHCs).

\begin{table}
\centering
\caption{
Results of fitting the periodograms derived from the four energy
bands with the broken power-law model (breaking from a slope of
$\alpha_{\rm low} = 1$ at low frequencies to $\alpha_{\rm hi}$). 
Errors on the model parameters correspond
to a 90 per cent confidence level for one interesting parameter
(i.e. $\Delta \chi^{2}=2.7$).
\label{tab:e-psd}}
\begin{center}
\begin{tabular}{lccc}                
\hline
          & Slope          & $f_{\rm br}$ &               \\
Band      & $\alpha_{\rm hi}$       & ($10^{-4}$ Hz)           & $\chi^{2}/dof$  \\
\hline
full        & $2.8_{-0.4}^{+0.2}$ & $5_{-3}^{+1}$    &  $43.5/32$   \\
soft        & $2.6\pm0.4 $        & $3_{-2}^{+3}$    &  $44.9/32$  \\
medium      & $2.7\pm0.5$         & $5\pm3$          &  $51.1/32$   \\
hard        & $2.3\pm0.4$         & $5\pm3$          &  $44.0/32$  \\
\hline
\end{tabular}
\end{center}
\end{table}

\section{Cross spectral analysis}
\label{sect:cross}

The cross spectrum is related to the PSD (sometimes known as the
auto spectrum). It offers a comparison of two simultaneous light
curves (e.g. from different energy bands) as a function of temporal
frequency. The cross spectrum can be used to derive  the coherence 
($\gamma^{2}(f)$) of
two light curves (a measure of how well correlated the variations in
the two light curves are) and time delays between the light
curves (see Vaughan \& Nowak 1997 and Nowak \et 1999 for an
explanation of these properties in 
terms of GBHC observations and Papadakis, Nandra \& Kazanas 2001 and
Vaughan \et 2003 for an application to Seyfert
galaxies). 

The coherence of \mrk\ was measured using the recipe given by Vaughan
\& Nowak (1997). The periodograms of the two light curves and their
cross periodogram were computed and binned by $N=15$ consecutive
frequencies per bin
%\footnote{The analysis was also performed using $N=40$ binning and the
%results were consistent.}. 
The coherence and its uncertainty were then
calculated using equation 8 of Vaughan \& Nowak (1997).  (Vaughan \et
2003 have shown that uncertainty estimates calculated in this way are
accurate for data  very similar to those examined here.) 

\begin{figure}
\rotatebox{270}{
\resizebox{!}{\columnwidth}{
{\includegraphics{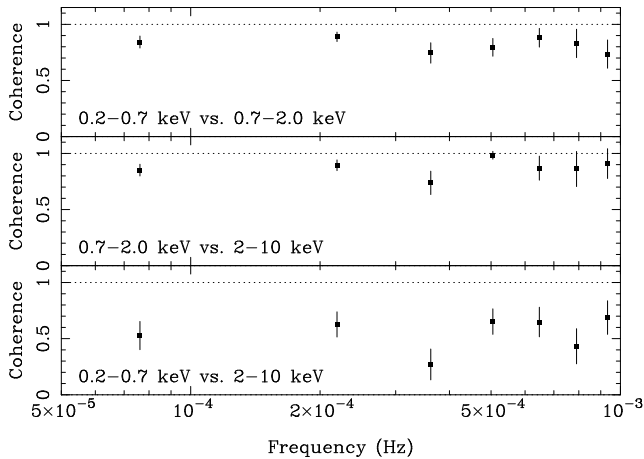}}}}
\caption{
Coherence function between light curves in the three energy bands.
The coherence (at a given frequency) decreases as the separation between
energy bands increases. At frequencies above $10^{-3}$~Hz the
Poisson noise power becomes comparable to the source power and the
coherence becomes difficult to measure (Vaughan \& Nowak 1997).
\label{fig:coherence}
}
\end{figure}

The resulting coherence functions (Fig.~\ref{fig:coherence})
show that, particularly between the soft and hard bands, the 
coherence is significantly below unity at all frequencies.
This means that either the relation between soft and hard bands
is non-linear or that a fraction ($1-\gamma^{2}(f)$) of the
hard band variance cannot be accounted for by variations
in the soft band. The
apparently poor (linear) correlation between variations in different bands is
confirmed by an analysis of the cross correlation function, estimated
using the Discrete Correlation Function (DCF; Edelson \& Krolik
1988). The DCF for the soft versus hard band light curves reaches a
peak (at zero time delay) of only 0.65. It should be noted that the
reduced coherence is not an artifact of Poisson noise in the data as
the recipe of Vaughan \& Nowak (1997) correctly accounts for this.
The low coherence means it is not possible to reliably search for
time delays between the different bands.

\begin{figure}
\rotatebox{270}{
\resizebox{!}{\columnwidth}{
{\includegraphics{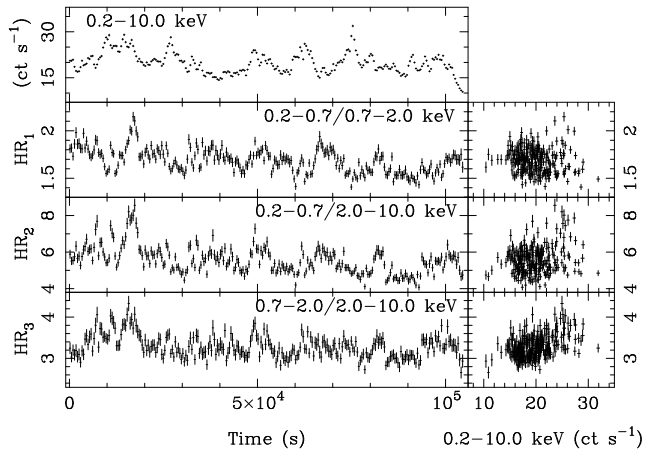}}}}
\caption{
Hardness ratios (in 400~s time bins). Top panel: the full band light curve
Lower panels: (left) hardness ratios plotted against time
and (right) against full band count rate.
\label{fig:hr}
}
\end{figure}

The hardness ratios were calculated to further examine the
relationship between variations in different energy sub-bands (Fig.~\ref{fig:hr}). 
The hardness ratios are significantly variable but show little
correlation with the total source flux. There is a slight trend for
the 0.7--2.0/2.0--10.0 keV ratio to soften as the source brightens
(as has been seen in other Seyfert 1s) but the ratios show a great
deal of scatter (much larger than expected from the size of the error bars) 
not obviously associated with changes in the full band
flux. This complements the coherence analysis and shows the
the relationships between energy sub-bands are complex.

\section{Discussion}
\label{sect:disco}

The shape of the high frequency power spectrum of \mrk\ is  consistent
with a power-law that breaks from a flat slope ($\alpha_{\rm low}
\approx 1$) to a steep slope ($\alpha_{\rm hi} \gsim 2$) at a
frequency $\sim 5 \times 10^{-4}$~Hz. There is no evidence for
discrete features such as quasi-periodic oscillations (previously
claimed by Boller \et 2001, but see also Benlloch \et 2001). Similar
broken power-law PSDs have been observed in other Seyfert 1s (Uttley
\et 2002; Vaughan \et 2003; Markowitz \et 2003; M$^{\rm c}$Hardy \et
2003). This shape for the high frequency PSD is similar to that of
GBHCs (Nowak \et 1999; Lin \et 2000). Not only is the shape 
the same but the amplitudes are similar. The power 
(in $f \mathcal{P}(f)$ units) measured at the break frequency is
$\sim 0.01$ in the 2--10~keV band, which compares with the power
at the high frequency break in Cyg X-1 of $\sim 0.02$
(Belloni \& Hasinger 1990). This strongly supports the idea
that the variability mechanisms, and perhaps the X-ray emission
processes in general, are the same in AGN and GBHCs.

In the best studied GBHC, Cygnus X-1 in its low/hard state, the PSD
breaks from $\alpha \approx 1$ to a steeper slope at $\sim 3$~Hz
(Belloni \& Hasinger 1990; Nowak \et 1999).  In its high/soft state
the position of the high frequency break increases to $\sim 10$~Hz
(Churazov \& Gilfanov \& Revnivtsev 2001).  It is not clear which of
these states is closer to the behaviour of Seyferts.  The
corresponding break frequency in \mrk\ is a factor of $\sim 6 \times
10^{3}$ or $\sim 2 \times 10^{4}$ times higher than that seen in Cyg
X-1, depending on whether the low/hard or high/soft state is used for
comparison.  Assuming a $10$~\Msun mass black hole in Cyg X-1 (Hererro
\et 1995), and further assuming that the characteristic break
timescale scales linearly with black hole mass (see Markowitz \et
2003) yields estimates for the mass of the black hole in \mrk\ of
$\sim 6 \times 10^{4}$ or $\sim 2 \times 10^{5}$~\Msun (comparing with
low/hard and high/soft state PSDs respectively). These estimates are
unusually low compared to the masses expected for Seyfert nuclei
(e.g. Wandel, Peterson \& Malkan 1999). Indeed, Wandel (2002)
estimated the mass of \mrk\ to be $\sim 10^{7}$~\Msun\ based on its
luminosity and optical broad-line properties. 

The highest mass estimate consistent with a linear scaling of
timescales, obtained using the 90 per cent lower limit on the  break
frequency in \mrk\ and scaling by the break seen in the high/soft
state in Cyg X-1, is $\sim 5 \times 10^{5}$~\Msun. The (unabsorbed
$0.2 - 10$~keV) X-ray luminosity of \mrk\ is $\sim 3 \times
10^{43}$~erg s$^{-1}$ (assuming $ H_0 = 70 $~km s$^{-1}$ Mpc$^{-1}$
and $ q_0 = 0.5 $).  Taking a fairly conservative estimate of the
bolometric luminosity of $10^{44}$~erg s$^{-1}$ and the highest black
hole mass estimate suggests \mrk\ is radiating at $\sim 1.5 L_{\rm
Edd}$. Of course, this depends on there being a linear scaling between
the break timescales in Cyg X-1 and those of Seyfert 1s. If this is
the case the implication seems to be that \mrk\ is accreting above the
Eddington limit.  

A further point is that if the slope of the low frequency PSD is
significantly different from the assumed $\alpha_{\rm low}=1$ then the
break frequency could be lower (fitting a model assuming $\alpha_{\rm
low}=0$ gave $f_{\rm br} = 1.6\pm0.8 \times 10^{-4}$~Hz). 
Comparing this to the low-frequency break in Cyg X-1
(the break from $\alpha \approx 0$ to $\alpha \approx 1$  in the
low/hard state) gives an even lower black hole mass estimate (see the
discussion in Uttley \et 2002). However, the fact that the high
frequency PSD is steep ($\approx 2.3$ in the hardest band) suggests 
a slope of $\approx 1$ below the break is more likely, by comparison with
the PSD of Cyg X-1. 

It is perhaps worth noting that \mrk\ has previously been classified a
Narrow-line Seyfert 1 (NLS1), a class of AGN often suggested to  be
accreting close to the Eddington limit (Boller, Brandt \& Fink
1996).  The PSDs of other NLS1s also seem to show break frequencies
somewhat higher than found in ``normal'' Seyfert 1s (Ark 564: Pounds \et
2001; Papadakis \et 2002; NGC 4051: M$^{\rm c}$Hardy \et 2003).  This
is consistent with the idea that NLS1s harbour lower mass black holes
accreting at a higher rate (relative to Eddington) compared to the
normal Seyfert population. 

It remains plausible that the mass-timescale relation is not perfectly
linear. The most important physical timescales associated with
accretion flows (e.g. section 5.8 of Frank, King \& Raine 1985) derive
from a distance and a speed over which some physical process operates
(such as viscosity). If the salient distance (such as the radius of
the innermost edge of the accretion disc) is approximately the same in
units of the gravitational radius ($r_{\rm g} = GM/c^2$) between Cyg
X-1 and AGN then the timescales should (to first order) scale linearly
with black hole  mass. However, if the relevant distance is not
constant then this linear scaling can break down. For example, the
innermost stable circular orbit around a black hole depends on its
dimensionless spin parameter ($a/M$), with  a spinning black hole
allowing a steady accretion disc to extend to smaller radii (in
gravitational units). If \mrk\ has a higher spin parameter than Cyg
X-1 one would perhaps expect the accretion disc to extend to smaller
$r_{\rm g}$ and the physical timescales to be somewhat smaller than
the linear expectation.  Such a situation would allow the black hole
mass of \mrk\ to be higher by a factor of a few (compared to the
linear assumption) and thereby allow for a sub-Eddington accretion
rate.  An independent estimate of the mass of \mrk\ (from
e.g. reverberation mapping or gas kinematic studies) would help
clarify this issue. 

The variations in different energy bands show a coherence   well below
unity at all frequencies examined. A loss of  coherence at high
frequencies has also been observed in \mcg\ (Vaughan \et 2003), but at
the lowest frequencies probed the coherence was much higher than
observed in \mrk.  On longer timescales the variations in NGC~7469
were consistent with unity coherence (Papadakis \et 2001).  In Cyg X-1
the coherence is known to be high (near unity) over a wide range in
frequencies but falls significantly below unity at high frequencies
(Nowak \et 1999). The reduced coherence means the relationship between
variations in different energy bands cannot be described by a single
(linear) transfer function.

If the X-rays are produced by inverse-Compton scattering in  a hot
corona, as is often thought (e.g. Haardt \& Maraschi 1991), then the
low coherence implies rapid changes in the coronal properties and/or
multiple independent emission sites.  As discussed by Nowak \& Vaughan
(1996), a static Comptonising corona should produce unity coherence.
Rapid changes in the properties (e.g. size, temperature, optical
depth) of the X-ray emitting region could cause changes in the
transfer function (relating soft to hard flux) on the timescales
probed and so reduce the coherence.  Alternatively the low coherence
may mean there is more than one transfer function operating
simultaneously, which could be physically realised if the X-ray
emission is produced at several  independent sites.

\xmm\ has opened the door on high frequency timing studies of AGN, a
door that has remained largely closed since the end of the \exosat\
mission. Future work must establish whether the high frequency power
and cross spectral properties observed in the handful of Seyfert 1s
with \xmm\ long-looks (see also Vaughan \et 2003 and M$^{\rm c}$Hardy
\et 2003) are universal or whether different objects show different
characteristics (perhaps as GBHCs show distinct states).

\section*{Acknowledgements}

Based on observations obtained with \xmm, an ESA science mission with
instruments and contributions directly funded by ESA Member States and
the USA (NASA). 
We thank Phil Uttley for many insightful discussions
on the details of PSD estimation and Iossif Papadakis for a swift
and useful referee's report.
SV thanks PPARC for financial support. 
ACF thanks the Royal Society for support.

\bsp
\label{lastpage}
\end{document}